\documentclass[lettersize,journal]{IEEEtran}

\usepackage{cite}
\usepackage[numbers]{natbib}
\usepackage{amsmath,amssymb,amsfonts}
\usepackage{algorithmic}
\usepackage{graphicx}
\usepackage{textcomp}
\usepackage{xcolor}
\usepackage{wrapfig}
\usepackage{hyperref}

\usepackage{verbatim}
\usepackage{subcaption}
\usepackage{array}
\usepackage[export]{adjustbox}

\usepackage{algorithm}
\usepackage{array}
\usepackage{textcomp}
\usepackage{stfloats}
\usepackage{url}
\usepackage{verbatim}
\usepackage{float}
\usepackage[affil-it]{authblk}
\title{Evolution, Challenges, and Optimization in Computer Architecture: The Role of Reconfigurable Systems}

\author[1]{Jefferson Ederhion}
\author[2]{Festus Zindozin}
\author[1]{Hillary Owusu}
\author[2]{Chukwurimazu Ozoemezim}
\author[3]{Mmeri Okere}
\author[4]{Opeyemi Owolabi}
\author[5]{Olalekan Fagbo}
\author[6]{Oyetubo Oluwatosin}

\affil[1]{University of Maryland, College Park, USA}
\affil[2]{University of Florida, USA}
\affil[3]{Madonna University, Nigeria}
\affil[4]{University of Bradford, United Kingdom}
\affil[5]{Ball State University, USA}
\affil[6]{Austin Peay State University, USA}

\begin{document}

\maketitle

\begin{abstract}
The evolution of computer architecture has led to a paradigm shift from traditional single-core processors to multi-core and domain-specific architectures that address the increasing demands of modern computational workloads. This paper provides a comprehensive study of this evolution, highlighting the challenges and key advancements in the transition from single-core to multi-core processors. It also examines state-of-the-art hardware accelerators, including Tensor Processing Units (TPUs) and their derivatives, RipTide and the Catapult fabric, and evaluates their strategies for optimizing critical performance metrics such as energy consumption, latency, and flexibility. 

Ultimately, this study emphasizes the role of reconfigurable systems in overcoming current architectural challenges and driving future advancements in computational efficiency.

\end{abstract}

\section{Introduction}

This paper explores the evolving landscape of computer architecture in the post-Moore's Law era, highlighting challenges posed by the power wall and the end of Dennard scaling. It describes the transition from single-core to multicore architectures, which provided a temporary solution but introduced complexities in parallel computing. 

With the advent of domain-specific architectures (DSAs), specialized accelerators such as the TPU have emerged, optimizing for specific workloads such as machine learning. However, as the demand for sparse matrix computations increases, variants such as Sparse-TPU and FlexTPU offer more efficient solutions. The paper also discusses RipTide, a flexible architecture designed to balance programmability and energy efficiency, and Catapult, which leverages FPGAs for datacenter flexibility. The study concludes by emphasizing the importance of innovative and flexible architectures that cater to diverse computational needs while optimizing energy consumption.

Section II provides an in-depth analysis of the evolution of computer architecture, highlighting the challenges posed by Moore's law, Dennard scaling, and the 'power wall'. It examines the transition from single-core to multi-core processors, and subsequent issues such as 'dark silicon' and the 'memory wall'. Additionally, it looks at various computing paradigms, such as heterogeneous computing and domain-specific architectures, which aim to maximize performance for specific tasks while reducing energy consumption.

Section III focuses on parallelism in computer architecture. It discusses instruction-level parallelism (ILP), data-level parallelism (DLP), and thread-level parallelism (TLP), each with their unique techniques and challenges. This section also compares various computing models, including the Von Neumann model, dataflow models, and hybrid models.

Sections IV through X explore state-of-the-art accelerators, including TPU, STPU, FlexTPU, and RipTide. These sections discuss how these accelerators are optimized for different performance metrics, such as energy consumption, flexibility, and latency.

\section{Evolution of computer architecture}
\subsection{Moore's Law, Dennard Scaling and the Bottleneck}

The history of computer architecture is deeply tied to Moore's law, formulated by Gordon Moore in 1965. Moore observed that the number of transistors on a microchip roughly doubles every two years. This exponential growth fueled miniaturization and advancements in integrated circuits, making them faster and more complex. Initially, chip designers focused on cramming more transistors into a single processor core to boost its clock speed, the rate at which it executes instructions. However, this approach soon hit physical limitations \cite{moore}.

As transistors shrunk, they became more susceptible to overheating and power consumption skyrocketed. This phenomenon is known as the power wall \cite{powerwall}. Compounding this issue was the end of Dennard scaling, another principle observed in the early days of microchips. Dennard scaling predicted that even as transistors miniaturized, their power consumption would remain relatively constant due to reduced voltage requirements. Unfortunately, this beneficial effect diminished as the transistors reached the atomic scales, making it impractical to further reduce the voltage without compromising performance \cite{moore}\cite{powerwall}.

\subsection{The Multicore Solution and Its Challenges}
The combined pressure from the power wall and the end of Dennard scaling forced a paradigm shift in processor design. Around the early 2000s, chipmakers began transitioning from single-core to multicore architectures. Multicore processors essentially pack multiple independent processing cores onto a single chip. This approach allowed parallel processing, where multiple tasks could be executed simultaneously, effectively improving overall performance without significantly increasing power consumption \cite{postdennard}.

However, the multicore revolution was not without its challenges. Software development had to adapt to efficiently utilize these multiple cores. Traditional software designed for single-core processors often could not leverage the parallelism offered by multi-core architecture. Furthermore, the increased number of cores introduced the concept of "dark silicon". Not all cores on a chip could be powered on simultaneously, due to limitations in heat dissipation and power delivery. This essentially rendered some transistors inactive, reducing overall efficiency \cite{darksilicon}\cite{room}.

In addition, there was the memory wall problem, which refers to the speed limitations of data transfer between the processor and the memory. This problem intensifies as the number of cores and the demand for memory bandwidth increase.

\begin{figure*}
\begin{center}
\makebox[\textwidth]{\includegraphics[width=\textwidth]{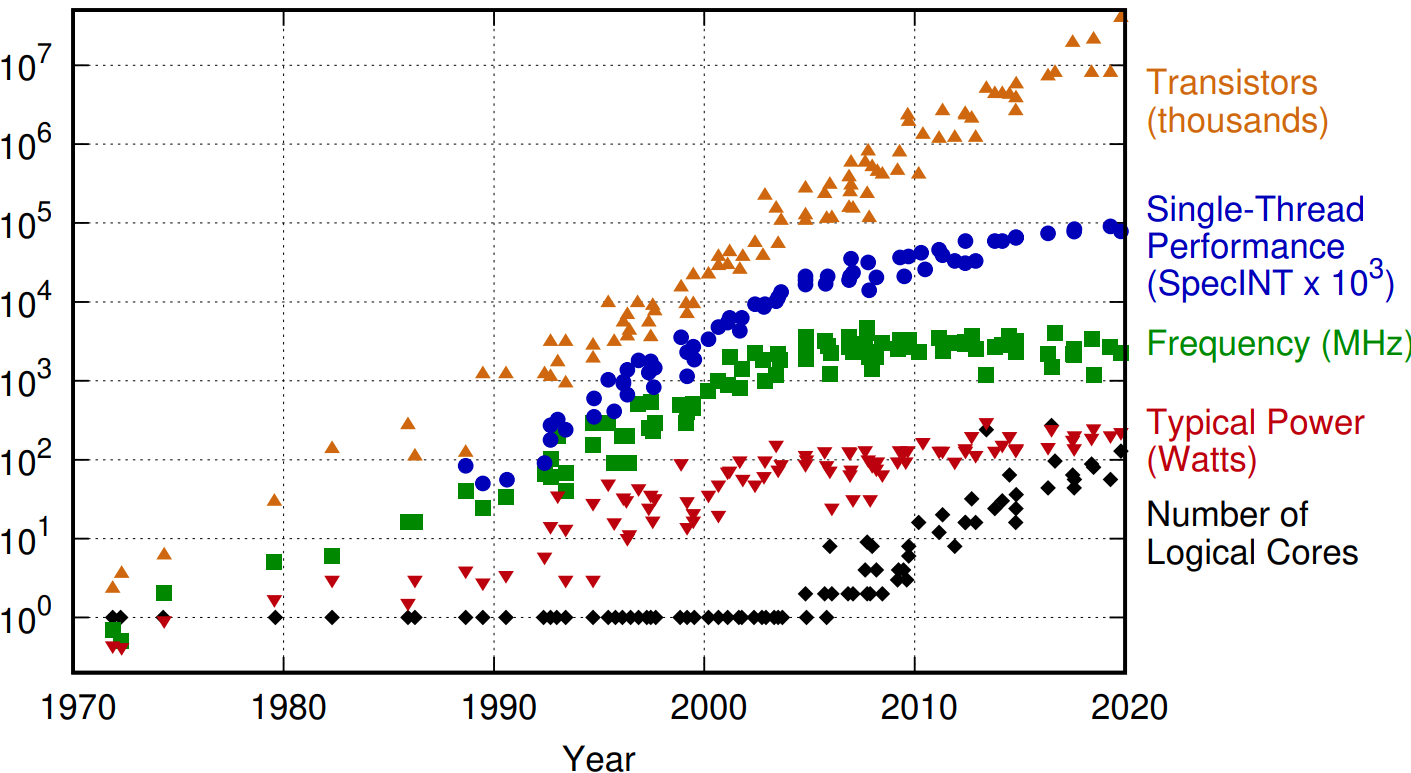}}
\caption{Trends in Microprocessor Technology from 1970 to 2020. Original data up to the year 2010 collected and plotted by M. Horowitz, F. Labonte, O. Shacham, K. Olukotun, L. Hammond, and C. Batten. New plot and data collected for 2010-2019 by K. Rupp}
\label{processor_trend}
\end{center}
\end{figure*}

Figure~\ref{processor_trend} illustrates the dramatic increase in transistor count on microprocessors over time, which is in accordance with Moore's law. This exponential growth fueled the advancements discussed previously, but ultimately led to the need for multicore architectures.

\subsection{Beyond Multicore: Heterogeneous Computing and Domain-Specific Architectures}

The pursuit of improved performance and energy efficiency continues to drive innovation in computer architecture. One such innovative approach is heterogeneous computing. In contrast to homogeneous computing, which exclusively relies on Central Processing Units (CPUs) for all tasks, heterogeneous computing combines different types of processors such as CPUs which handles general-purpose tasks while controlling the overall system operations, Graphics Processing Units (GPUs) which perform parallel processing tasks such as rendering and Machine Learning, and specialized accelerators on a single chip. Each type of processor in a heterogeneous computing architecture is optimized for specific tasks, leading to improved performance and reduced power consumption for workloads that can be parallelized across these diverse cores while also allowing efficient communication and resource sharing between all components\cite{room}.


These specialized accelerators are typically referred to as Domain Specific Architectures (DSAs). DSAs are custom-designed processors that are tailored for specific computational tasks, such as machine learning or video processing. They take advantage of the inherent characteristics of these domains to achieve optimal performance and energy efficiency. For example, the Eyeriss accelerator is specifically designed for Convolutional Neural Networks (CNNs), a type of artificial neural network commonly used in deep learning applications. Eyeriss optimizes data movement and minimizes energy consumption, which are critical aspects of deep learning algorithms \cite{eyeriss}.\\

When designing a DSA, computer architects must make two crucial decisions:

\begin{enumerate}
  \item Type of Parallelism: Parallelism is a fundamental concept in modern computer architecture aiming to improve performance by running multiple instructions or operations concurrently. Different types of parallelism exist, such as instruction-level parallelism, data-level parallelism, and thread-level parallelism. The architect needs to choose the type that best suits the target application to maximize efficiency.
  \item Computing Model: The computing model defines how data are processed and flow within the DSA. Common models include the von Neumann model and the dataflow model. The architect must select the model that aligns best with the use case and the chosen type of parallelism.
\end{enumerate}

The next sections will briefly discuss these concepts.

\section{Types of Parallelism in Computer Architecture}
There are three popular kinds of parallelism, which will be discussed below.

\subsection{Instruction Level Parallelism (ILP)}

ILP focuses on extracting parallelism within a single stream of instructions. It leverages the fact that some instructions in a program may be independent of each other, meaning they do not rely on the results of previous instructions and do not produce results that affect subsequent instructions. Identifying these independent instructions allows processors to execute them concurrently, improving overall program execution speed. 

Several techniques are used to exploit ILP:
\begin{itemize}
  \item Instruction pipelining: This technique overlaps the execution of different stages of an instruction (fetching, decoding, execution, memory access, and writing results) with those of subsequent instructions \cite{hennessy}.
  \item Multiple instruction issue: This technique is a feature of modern processors that allows them to decode and issue multiple instructions simultaneously, given that they are independent. The primary advantage of this approach is that it enables the processor to keep its execution units active and reduce idle time, thus improving overall efficiency \cite{hennessy}.
  \item Out-of-order execution: It involves the analysis of instructions, their reordering for efficient execution, and their execution outside of their original program order. This technique helps to hide memory access latencies. A popular algorithm used for out-of-order execution is the Tomasulo Algorithm \cite{hennessy}
\end{itemize}

However, ILP can be limited by dependencies which include data dependencies, control dependencies, and structural dependencies.  

\begin{figure}[htbp]
\centering
\includegraphics[width=\columnwidth]{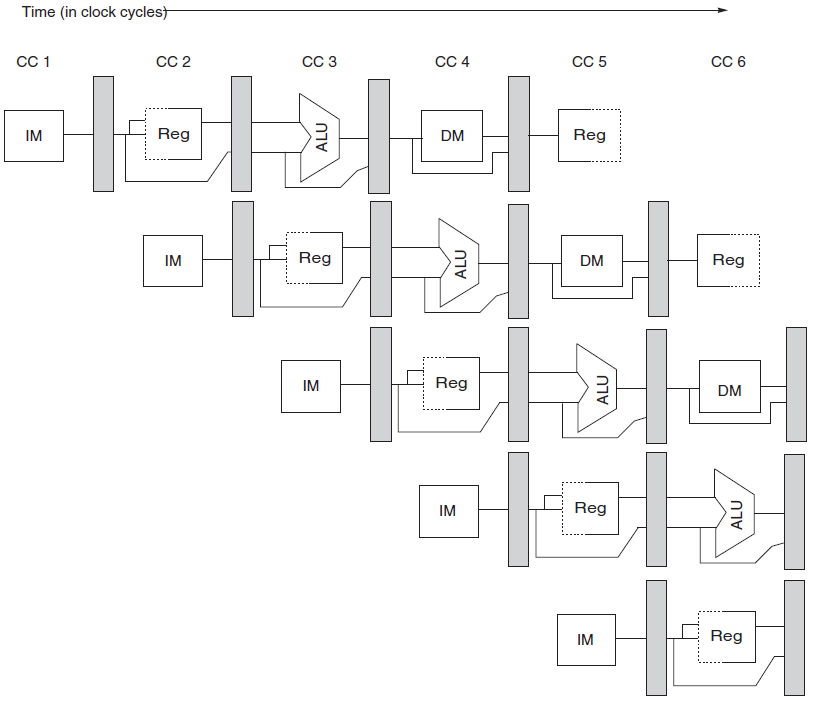}
\caption{Instruction Pipelining \cite{hennessy}}
\label{pipeline}
\end{figure}

Figure~\ref{pipeline} illustrates an example of ILP using instruction pipelining

\subsection{Data Level Parallelism (DLP)}

DLP involves the simultaneous execution of identical operations across multiple data elements, thereby significantly improving computational efficiency. This form of parallelism is leveraged through vector processing, Single Instruction Multiple Data (SIMD) architectures, and Graphics Processing Units (GPUs). DLP is particularly advantageous in applications such as scientific computing, image processing, and artificial intelligence, where large datasets require repeated application of similar operations.

By concurrently processing multiple data elements, DLP considerably improves throughput for data-intensive tasks, making it an indispensable technique in high-performance computing environments. This is particularly evident in applications that involve repetitive and computationally intensive operations, such as matrix multiplications or convolution operations in neural networks.

A common instance of DLP can be observed in everyday programming. Consider a simple "for loop" written in C, as shown in Figure~\ref{for}. When compiled with the GCC compiler using the highest optimization level, the generated assembly code (Figure~\ref{assembly}) reveals how the processor automatically translates the loop into SIMD instructions, leveraging Intel's Streaming SIMD Extensions (SSE). This process, often abstracted by the compiler, demonstrates how modern compilers and hardware architectures automatically harness DLP to enhance performance, even in relatively simple code.

\begin{figure}[h]
\centering
\begin{subfigure}{\columnwidth}
    \begin{verbatim}
    for(i = 0; i < SIZE; i++) 
    {
        a[i] = i * i + 5;
    }    
    \end{verbatim}
    \caption{}
    \label{for}
\end{subfigure}
    
\begin{subfigure}{\columnwidth}
    \begin{verbatim}
    .L8:
        movdqa  %xmm2, %xmm1
    .L3:
        movdqa  %xmm1, %xmm5
        movdqa  %xmm1, %xmm2
        addq    $16, %rax
        pmuludq %xmm1, %xmm5
        psrlq   $32, %xmm1
        pshufd  $8, %xmm5, %xmm0
        pmuludq %xmm1, %xmm1
        pshufd  $8, %xmm1, %xmm1
        paddd   %xmm4, %xmm2
        punpckldq       %xmm1, %xmm0
        paddd   %xmm3, %xmm0
        movdqa  %xmm0, -16(%rax)
        cmpq    %rdx, %rax
        jne     .L8
    \end{verbatim}
    \caption{}
    \label{assembly}
\end{subfigure}
\caption{Demonstration of DLP. (a) Sample 'for loop' written in C. (b) Assembly code generated by the GCC compiler, highlighting the use of SIMD instructions via Intel's Streaming SIMD Extensions (SSE), which efficiently parallelizes data processing.}
\label{dlp}
\end{figure}

\subsection{Thread Level Parallelism (TLP)}
TLP enhances computational throughput by concurrently executing multiple threads. Threads may represent independent tasks or subtasks of a larger application, allowing multi-threaded programs to distribute workloads effectively across multiple cores or processors. This concurrent execution enables TLP to fully exploit the available computing resources, leading to significant performance gains, especially in multicore processors and distributed systems \cite{hennessy}\cite{hybrid}.

TLP is crucial for maximizing the throughput of multi-threaded applications by leveraging the power of modern hardware architectures. It enables applications to scale horizontally as processing power increases, making it a key factor in improving performance and efficiency. This parallelism is essential for workloads that benefit from simultaneous execution of multiple threads, such as web servers, databases, and scientific simulations \cite{hybrid}.

However, TLP also presents challenges. Synchronization overhead can arise from managing data shared across threads, potentially leading to bottlenecks. Load balancing is another critical aspect, requiring careful design to ensure that no single thread is overloaded while others remain underutilized. Furthermore, the increased complexity in software design demands more intricate programming techniques to achieve optimal performance without introducing errors.

\section{Computing Models}

Computer architects must carefully select the computing model that best aligns with their use case when designing DSAs. There are two primary computing models: the Von Neumann model and the Dataflow model.

\subsection{Von Neumann Model}
The Von Neumann execution model, a classic architecture, fetches program instructions from memory, decodes them, executes them, and stores the results. The program counter (PC) determines the next instruction to be executed, facilitating sequential execution. Operands are retrieved from a centralized memory or registers, forming the foundation of traditional computer architectures. This process drives a linear instruction flow in sequential programming \cite{hennessy}\cite{hybrid}. The Von Neumann architecture, offering high flexibility, is essential in most general-purpose processors.

Due to the inherent sequential execution of the model, computations tend to be slow. To enhance performance, various types of parallelism can be utilized. For instance, Instruction Level Parallelism (ILP) can be exploited through techniques such as pipelining and superscalar architectures. Data-Level Parallelism (DLP), on the other hand, can be implemented using Single Instruction, Multiple Data (SIMD) extensions. Additionally, multi-threading can be used to improve throughput. \cite{hybrid}.

\begin{figure}[htbp]
\centering
\includegraphics[width=\columnwidth]{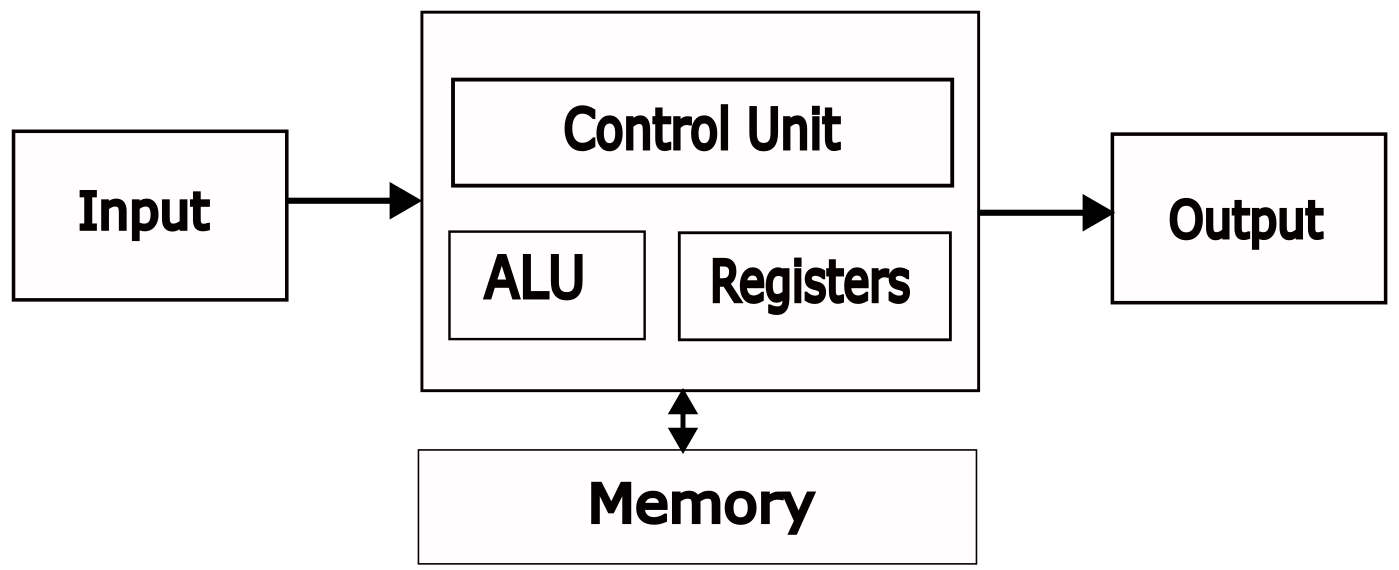}
\caption{Von Neumann Computing Model}
\label{neumann}
\end{figure}

In Figure~\ref{neumann}, we present an example of the Von Neumann computing model, showing its structure and the sequential instruction execution characteristic of this model.

\subsection{DataFlow Model}
In the dataflow model, the execution of instructions is not based on a predefined sequence as in the Von Neumann model. Instead, instructions are executed as soon as their input operands become available. This approach eliminates the need for a program counter. The model uses a data dependency graph, where each instruction directly passes its output to the subsequent instructions that require its results. This structure creates an efficient flow of data through the program, allowing a more effective execution of independent instructions. In this way, the dataflow model bypasses the need for a centralized register file, enhancing the efficiency of instruction execution \cite{hybrid}.

\begin{figure}[htbp]
\centering
\includegraphics[width=0.5\columnwidth]{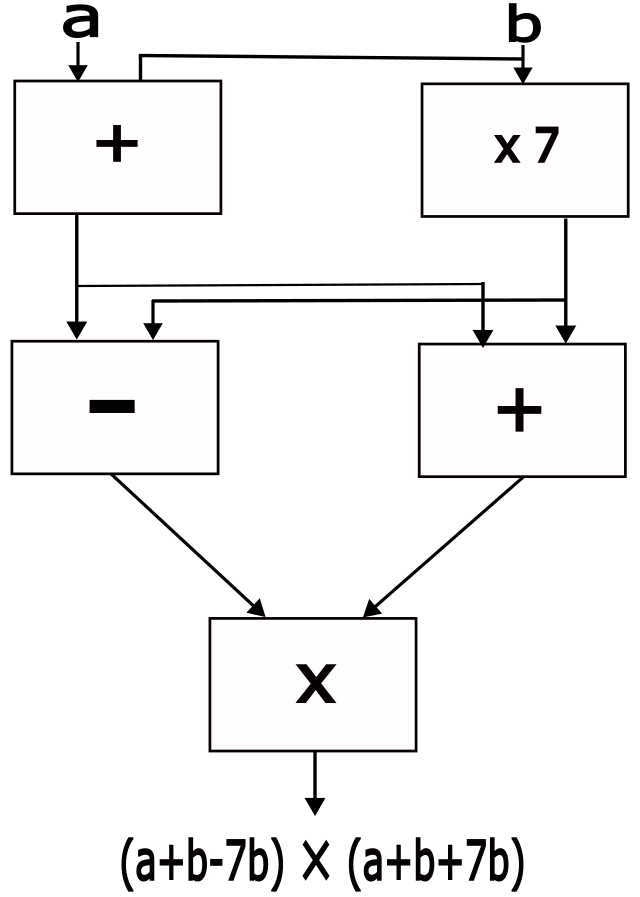}
\caption{A basic dataflow graph using an adder tree configuration \cite{onur}}
\label{addertree}
\end{figure}

Figure~\ref{addertree} illustrates a basic dataflow graph for the mathematical expression $$ (a + b - 7b) \times (a + b + 7b) $$. The model uses an adder tree configuration to efficiently execute the mathematical operation with two processing elements (PE). Assuming that each operation takes one cycle, the adder tree completes the entire operation in three cycles, compared to five cycles for sequential execution. To optimize performance, the mathematical expression can be rearranged so that operations with the same latency execute in the same cycle.

\begin{figure}[htbp]
\centering
\includegraphics[width=\columnwidth]{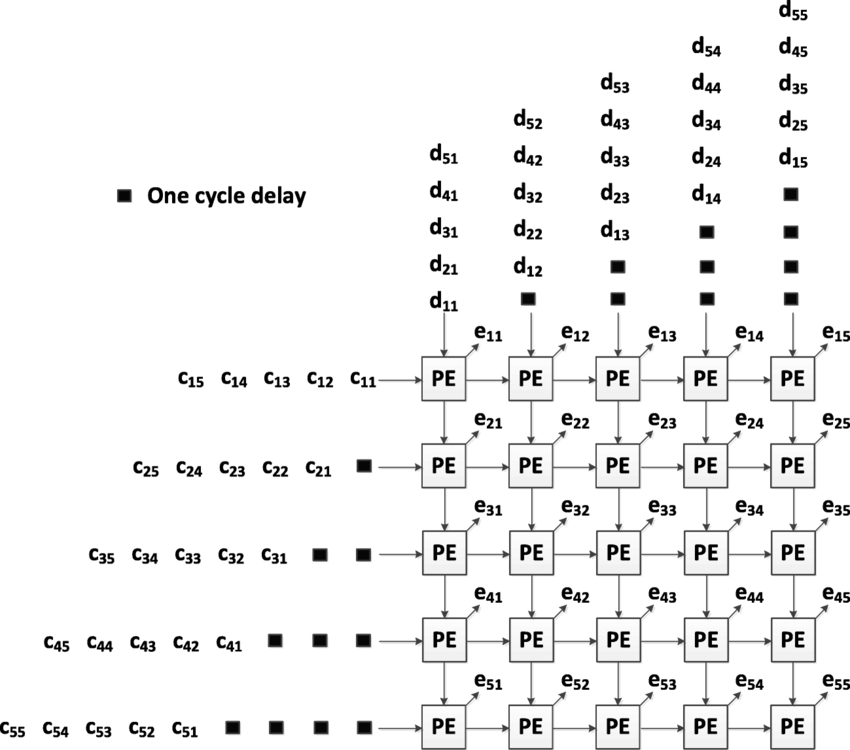}
\caption{Systolic array \cite{systolicref}}
\label{systolicarray}
\end{figure}

Systolic arrays are often favored over adder trees for matrix multiplication because of their efficiency and their ability to handle high data reuse rates, a key requirement in matrix multiplication. A systolic array is composed of a grid of processing elements (PEs), typically Multiply-Accumulate (MAC) units, which are closely interconnected \cite{meissa}. As shown in Figure~\ref{systolicarray}, a systolic array performing matrix-matrix multiplication allows both input data and results to flow through the grid. This is in contrast to adder trees, where only the results flow. This characteristic of systolic arrays enables more efficient data processing and promotes optimal data reuse.

The primary advantage of the dataflow model lies in its data reuse efficiency. By prioritizing data movements, dataflow architectures can significantly minimize unnecessary data transfers and memory accesses. These are often the leading causes of energy consumption and latency in traditional Von Neumann systems. In addition, the inherent flexibility of the dataflow model enables the architectures to be highly reconfigurable, allowing them to efficiently adapt to a variety of computational tasks and data types. This adaptability proves to be crucial when dealing with a diverse range of deep learning models and algorithms, each requiring different computational resources \cite{meissa}.

 \subsection{Hybrid}
The hybrid model combines the strengths of both the Von Neumann and dataflow models to take advantage of sequential execution and parallelism. By integrating the programmability and sequential execution of the Von Neumann model with the data-driven parallel execution of the dataflow model, the hybrid model seeks to balance ease of programming and resource utilization. This approach is particularly beneficial for applications that require sequential and parallel processing \cite{hybrid}.

\section{Choosing the Right Architecture for Domain-Specific Accelerators}
The choice of computing model and type of parallelism when designing a domain specific architecture (DSA) depends on the specific requirements of the accelerator. For example, if the accelerator is intended for a device that is used in remote locations and operates on battery power, such as energy harvesting devices, energy efficiency is paramount. This is because it is not feasible to frequently replace the battery in these devices. In such cases, an Application-Specific Integrated Circuit (ASIC) might be the optimal choice as it offers superior energy efficiency, being optimized for a specific use case. Additionally, if the goal is to optimize latency and/or throughput for a specific workload, an ASIC would also be the ideal choice. 

Currently, most ASICs used for neural network applications typically use the dataflow model because matrix operations are well suited for this model. However, there may be instances where the accelerator needs to perform multiple operations on the workload for which the dataflow model is not optimized. In such cases, using only the dataflow model would not be energy efficient. The hybrid model, which integrates the Von Neumann model for computations requiring sequential computation and the dataflow model for the rest of the accelerator, would be more suitable in this scenario.

It is common to want an accelerator to be able to perform computations on different types of workload. Ideally, an architecture that offers the best flexibility and energy efficiency would be the best hardware choice. However, achieving this performance requirement on any hardware is challenging. In this case, an ASIC would be the least suitable choice because of its lack of flexibility. The CPU, which would provide the best flexibility, is very energy inefficient. A compromise could be to use a reconfigurable architecture, which offers flexibility while also being energy efficient.

\begin{figure}[htbp]
\centering
\includegraphics[width=\columnwidth]{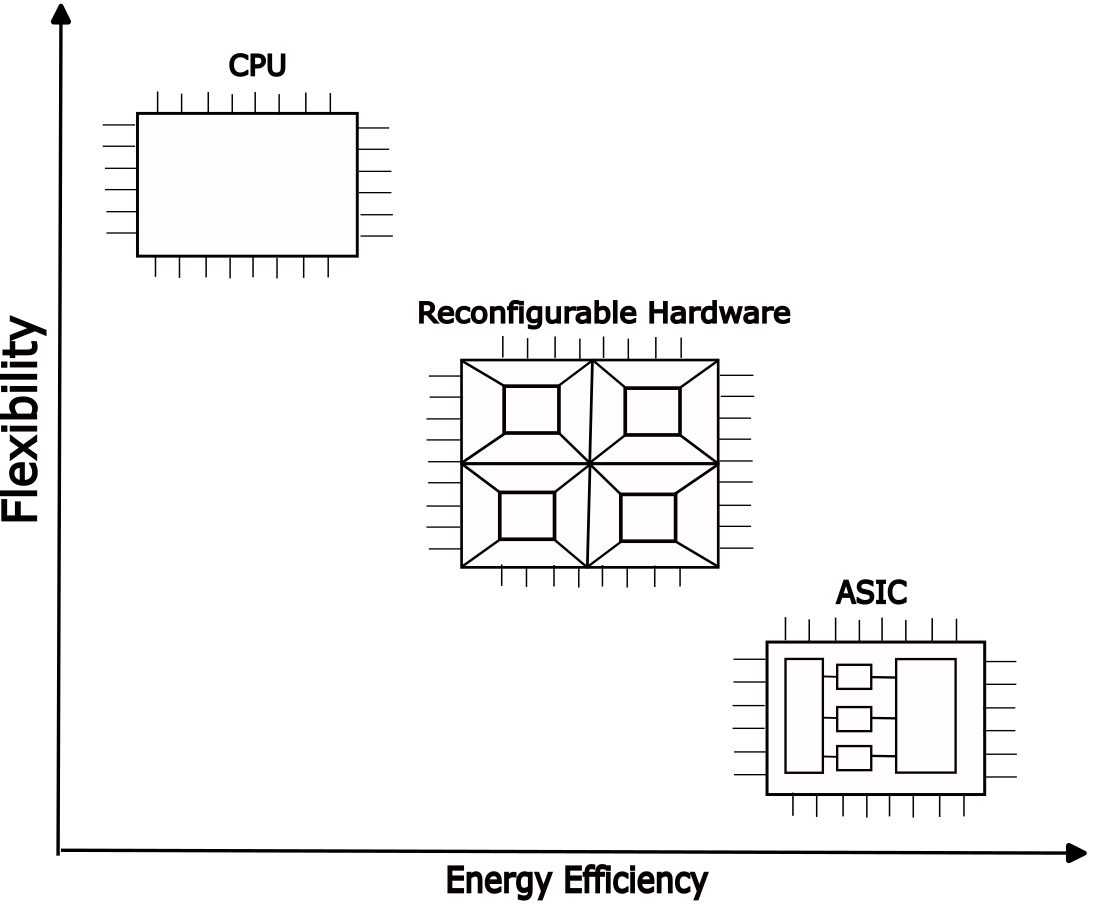}
\caption{Trade-off Between Flexibility and Energy Efficiency in Various Hardware Choices}
\label{flexibility/eenergy}
\end{figure}

Figure~\ref{flexibility/eenergy} illustrates the trade-off between flexibility and energy efficiency for various hardware choices.

Two popular reconfigurable architectures that are currently in use are the Coarse-Grained Reconfigurable Array (CGRA) and the Field-Programmable Gate Array (FPGA). The choice between these two largely depends on the specific goals of the accelerator. If the accelerator requires maximum flexibility, FPGAs are generally the best choice. However, in applications where there is a need to optimize latency and energy efficiency, CGRA tends to be the better option. This is primarily because CGRAs operate at a coarser granularity than FPGAs, leading to shorter reconfiguration times and higher energy efficiency. \\

In the following sections, we will discuss various popular DSA that have been designed with a focus on latency and/or throughput, energy efficiency, and flexibility. These DSAs include the Tensor Processing Unit (TPU) and its variants SparseTPU and FlexTPU. We will also look at RipTide, an accelerator designed for both energy efficiency and flexibility. Lastly, we will explore the use cases of reconfigurable hardware in data centers.

\section{Tensor Processing Unit}
This section discusses the Tensor Processing Unit (TPU) as presented in the paper "In-Datacenter Performance Analysis of a Tensor Processing Unit" by Norman P. Jouppi et al. \cite{TPU}.

Machine learning and artificial intelligence applications present several unique challenges that traditional hardware architectures struggle to handle efficiently. These challenges include:

\begin{enumerate}
  \item Computational demands: Machine learning models, especially neural networks, require substantial computational resources. The TPU is designed to efficiently handle these high computational demands.
  \item Power Efficiency: As the scale of data and complexity of the models increase, so does the power consumption. The TPU provides a high throughput per watt, making it a more energy-efficient solution compared to traditional CPUs and GPUs.
  \item Latency Requirements: Many real-world applications, such as autonomous driving or real-time translation, require low latency. The TPU deterministic execution model is better suited to the 99th percentile response time requirement of these applications.
  \item Memory Management: Machine learning models often require large amounts of memory to store intermediate results, weights, and biases. The TPU comes with a large, software-managed on-chip memory to efficiently handle these requirements.
  \item Scaling: As machine learning models become more complex and data sets grow larger, the need for hardware that can scale with these increases becomes apparent. TPUs are designed to work both independently and as part of a larger system, allowing easy scaling as computational needs grow.
\end{enumerate}

\subsection{TPU Architecture}
The TPU is an hardware accelerator designed specifically for machine learning workloads. At the heart of the TPU is a large matrix multiplication unit, which is capable of performing 65,536 8-bit multiply-and-add operations in a single cycle. 

It uses a systolic array architecture for its matrix multiplication unit. This design choice is based on the observation that a significant portion of computation in many machine learning workloads consists of matrix operations. The systolic array architecture allows for high computational efficiency and throughput by enabling multiple operations to be performed simultaneously. In addition, it minimizes the need for data movement, a major source of energy consumption in traditional architectures. 

\begin{figure}[h]
\centering
\begin{subfigure}{0.6\columnwidth}
    \centering
    \includegraphics[width=\columnwidth]{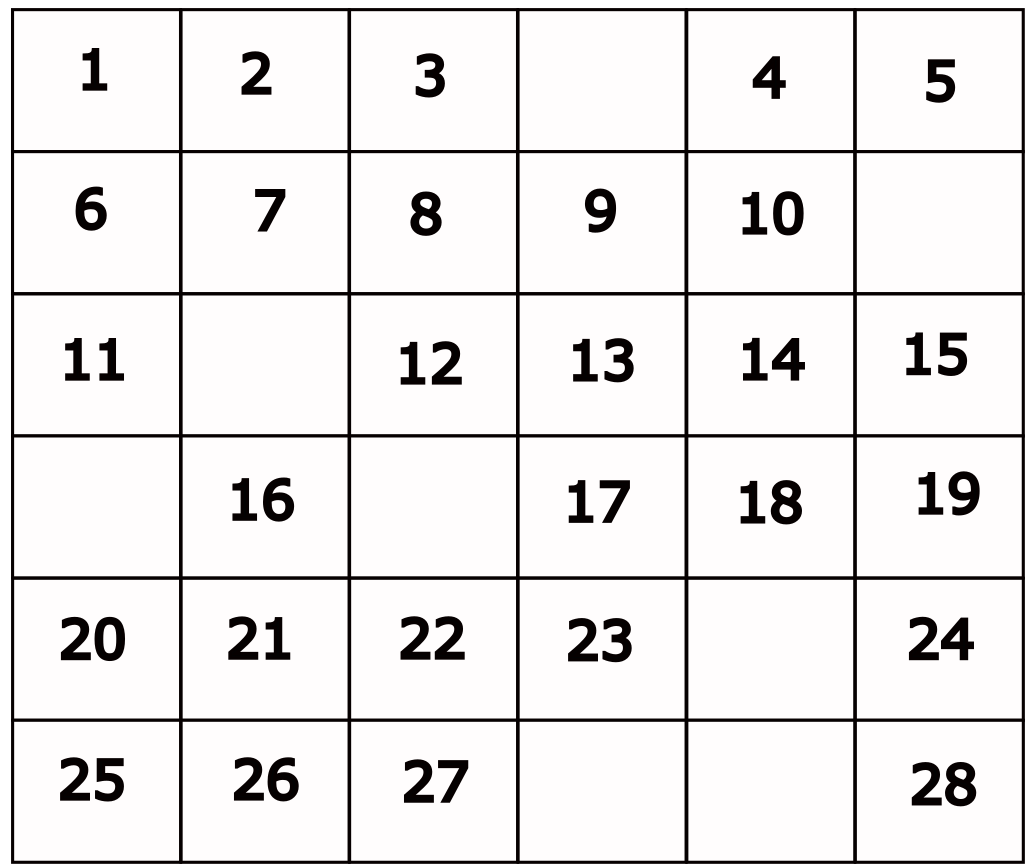}
    \caption{}
    \label{dense}
\end{subfigure}
    
\begin{subfigure}{\columnwidth}
    \centering
    \includegraphics[width=\columnwidth]{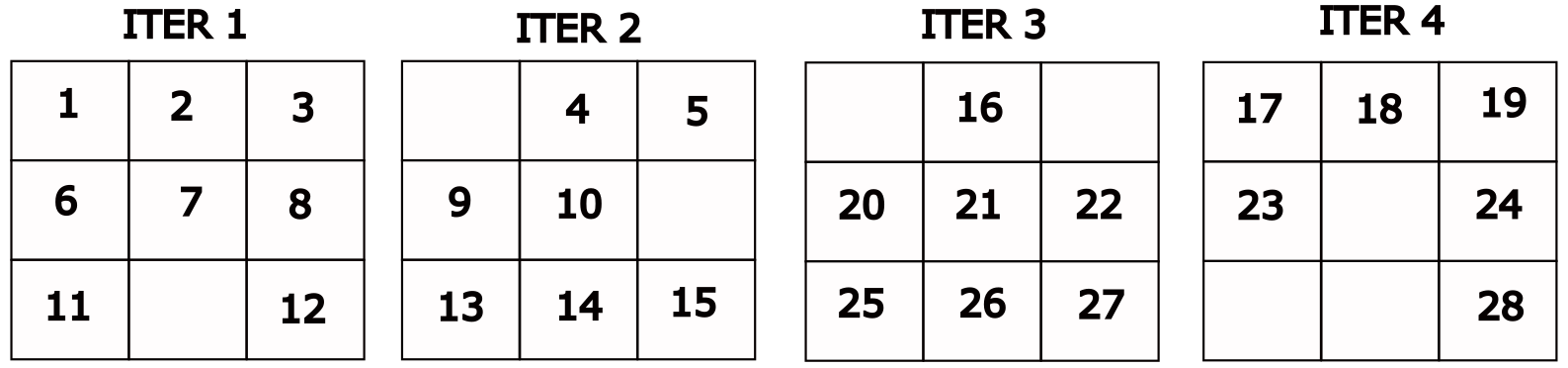}
    \caption{}
    \label{tpuiter}
\end{subfigure}
\caption{Mapping of (a) a dense matrix of size 6x6 to (b) 3x3 TPU systolic array \cite{flextpu}}
\label{TPUarr}
\end{figure}

One of the standout features of the TPU is its large, software-managed on-chip memory. This design allows for efficient data reuse, a common requirement in machine learning workloads. It also reduces the need for costly data transfers between the processor and off-chip memory, further enhancing the efficiency of the TPU architecture.

\subsection{Deterministic Execution}
In a TPU, operations are executed in a predictable and consistent manner. This is in contrast to CPUs and GPUs, which often employ dynamic scheduling and out-of-order execution to optimize performance. Although these techniques can improve average-case performance, they can also lead to variability in execution times, making it difficult to guarantee a specific response time.

On the other hand, TPUs are designed to perform a large number of operations simultaneously in a highly structured way, specifically matrix multiplications, which are at the heart of many machine learning workloads. Data flow through the TPU systolic array in a predictable pattern and the same operation is performed at each step. This means that for a given input size, the execution time is constant, regardless of the specific values of the input data.

This deterministic execution model allows TPUs to consistently meet the strict latency requirements of many machine learning applications, making them particularly well suited for real-time or near-real-time applications where predictability is key.

\subsection{Limitations of TPU}
TPUs are highly energy efficient when performing computations on dense matrices. Dense matrices are those with a high number of nonzero elements. An example of where a dense matrix is commonly found is in image processing, where the intensity of each pixel is represented by a numeric value between 0 and 255 \cite{verma2023sparse}. Operations on dense matrices are both memory and computationally intensive. As a result, it is common practice to prune a dense matrix to reduce the number of computations performed on it. This process introduces sparsity to the matrix.

Sparse matrices are matrices with a high number of zero elements. Sparsity in dense matrices can also be introduced through quantization. Furthermore, some applications, such as graph and recommendation systems, produce sparse matrices by default.

While TPUs are energy efficient for dense matrix computations, they are less so when used for sparse matrices, although they are still more energy efficient than CPUs and GPUs. The reason for this lies in the structure of the TPU itself. Specifically, the presence of sparsity in the matrices can lead to unutilized PEs within the TPU, thereby reducing its overall computational efficiency.

Figure~\ref{TPUarr} provides an illustration of how a 6x6 dense matrix is mapped to a 3x3 TPU systolic array. As shown in the figure, the TPU partitions the input matrix directly, based on the shape of the systolic array. This process requires four iterations for execution. It is important to note that even if the matrix were sparse, the TPU would still require four iterations for execution due to its method of partitioning the input matrix. This is because the TPU partitioning approach does not account for the sparsity of the matrix.\\

Several accelerators have been designed to handle sparse computations. This paper will discuss FlexTPU and STPU, accelerators that were designed by repurposing the TPU, thereby improving its energy efficiency. The idea behind repurposing the TPU is to minimize the number of unutilized PEs during computations.

\section{Sparse-TPU}
This section discusses the Sparse-TPU (STPU) as presented in the paper "Sparse-TPU: Adapting Systolic Arrays for Sparse Matrices" by Xin He et al. \cite{STPU}

The Sparse Tensor Processing Unit (STPU) addresses the challenge of efficiently handling sparse matrices. It recognizes that the traditional systolic array architecture, while highly efficient for dense matrices, can be adapted to handle sparse matrices as well.

In a traditional systolic array, such as that of TPUs, the PEs containing zeros in a sparse matrix are unutilized, leading to wasted computational resources and energy. The STPU addresses this inefficiency by repurposing these PEs.  

The STPU introduces a comprehensive framework that maps sparse data structures to a 2D systolic-based processor in a scalable manner. This allows the STPU to handle both dense and sparse matrices efficiently, thereby improving the utilization of the PEs and the overall energy efficiency. A key feature of the STPU is its ability to perform column merging when mapping the input matrix onto its systolic array.

\begin{figure}[h]
\centering
\begin{subfigure}{0.6\columnwidth}
    \centering
    \includegraphics[width=\columnwidth]{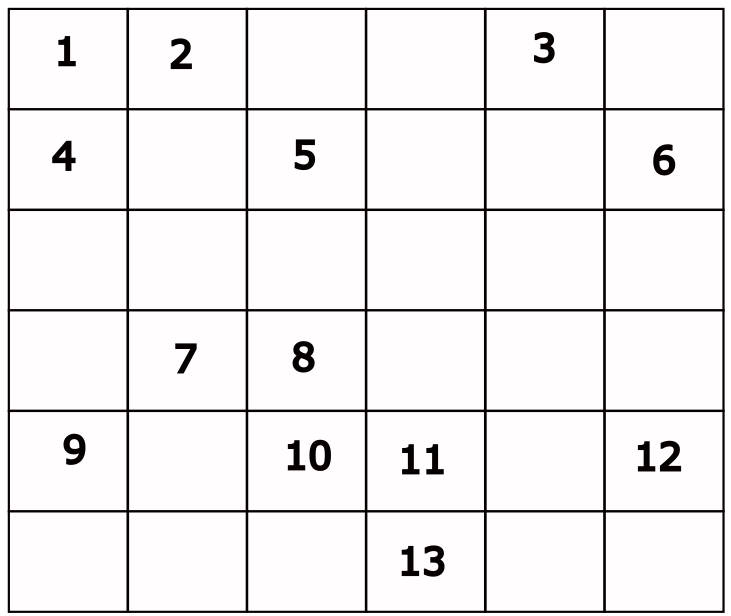}
    \caption{}
    \label{sparse}
\end{subfigure}
    
\begin{subfigure}{\columnwidth}
    \centering
    \includegraphics[width=\columnwidth]{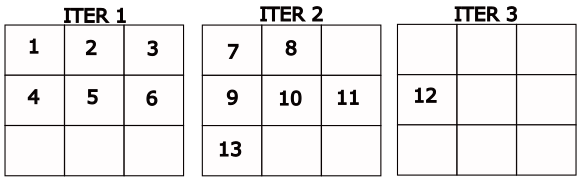}
    \caption{}
    \label{stpu_sys}
\end{subfigure}
\caption{Mapping of (a) a sparse matrix of size 6x6 to (b) 3x3 STPU systolic array \cite{flextpu}}
\label{stpu_array}
\end{figure}

Take, for example, Figure~\ref{stpu_array}, which shows how a 6x6 sparse matrix is mapped onto a 3x3 STPU systolic array. The STPU merges columns before mapping them onto the systolic array, enabling computations on the sparse matrix to be completed in just three iterations. Despite this design offering superior energy efficiency compared to TPU, there are instances where some PEs remain unutilized during computations. This occurs when there is a disproportionate number of non-zero columns across the rows, leading to an increase in unutilized PEs. For example, the value 12 of the sparse matrix required an additional iteration to map to the systolic array, despite the availability of slots in the systolic array in the previous iteration.

Additionally, the STPU handles both sparse matrix-vector (SpMV) and sparse matrix-matrix (SpMM) operations.

\section{FlexTPU}

This section discusses FlexTPU, as presented in the paper “Squaring the Circle: Executing Sparse Matrix Computations on FlexTPU - A TPU-like Processor” by Xin He et al. \cite{flextpu}.

The FlexTPU was designed by the same authors who developed the STPU. Although both TPU and STPU can execute SpMV and SpMM, the TPU is notably inefficient for these operations. The FlexTPU, on the other hand, was specifically designed to execute SpMV operations efficiently.

To handle these SpMV computations effectively, the FlexTPU employs a new mapping technique that minimizes the processing of zeros and maximizes the utilization of the TPU's systolic array. This approach enhances the flexibility of TPUs, enabling them to handle a wider range of computational tasks.

The FlexTPU achieves efficient mapping of sparse matrices onto the systolic array through a two-step process:

\begin{enumerate}
  \item Z-Shape Mapping: The first step involves a lightweight Z-shape mapping of sparse matrices onto the systolic array. This mapping aims to eliminate the processing of zeros as much as possible, regardless of the sparsity and non-zero distribution. The Z-shape mapping arranges the nonzero elements of the sparse matrix in a zigzag pattern across the systolic array, thereby maximizing the utilization of the PEs in the array.
  \item SpMV Dataflow Execution: Building on the mapping, an SpMV dataflow is executed by an array of PEs, which are slightly modified versions of the conventional TPU PE. This dataflow design ensures efficient computations following the Z-shaped mapping. Additionally, Z-shape mapping facilitates on-the-fly matrix condensing from the widely used compressed sparse matrix (e.g., CSR) representation. This is achieved by a proposed sparse data loader that includes an on-chip row decoder and parallel nonzero loaders.

\end{enumerate}

This combination of Z-shape mapping and SpMV dataflow execution, along with on-the-fly matrix condensing, enables FlexTPU to handle SpMV computations efficiently.

\begin{figure}[h]
\centering
\begin{subfigure}{0.6\columnwidth}
    \centering
    \includegraphics[width=\columnwidth]{sparse_matrix.png}
    \caption{}
    \label{sparse2}
\end{subfigure}
    
\begin{subfigure}{\columnwidth}
    \centering
    \includegraphics[width=0.7\columnwidth]{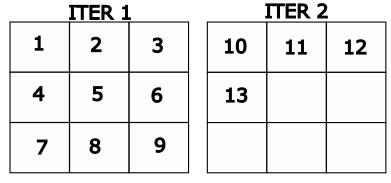}
    \caption{}
    \label{flex}
\end{subfigure}
\caption{Mapping of (a) a sparse matrix of size 6x6 to (b) 3x3 FlexTPU systolic array \cite{flextpu}}
\label{flextpu_array}
\end{figure}

Figure~\ref{flextpu_array} illustrates the mapping of a previously used 6x6 sparse matrix onto a 3x3 FlexTPU systolic array. As shown in the figure, the Z-shape mapping enables full utilization of the systolic array. This is achieved by consecutively mapping the non-zero values onto the systolic array, allowing the computation to be completed in just two iterations. This is more efficient compared to the three iterations required by the STPU. The difference in the number of iterations becomes even more significant when dealing with a matrix with a high degree of sparsity.

Unlike the STPU, the FlexTPU does not have unutilized PEs when used for SpMV. It achieves a 3.55× speedup and 3.27× energy saving when compared to the Sparse-TPU. This makes FlexTPU a more efficient and effective solution for handling SpMV computations.\\

\begin{table*}[] 
    \centering
    \caption{Comparison of TPU, STPU, and FlexTPU Architectures}
    \label{tab:comparison} 
    \small 
    \begin{tabular}{|c|p{2.5cm}|p{3.0cm}|p{3.0cm}|} 
    \hline
    \textbf{Features} & \textbf{TPU} & \textbf{STPU} & \textbf{FlexTPU} \\
    \hline
    Processing Elements & MAC units & MAC units & MAC units \\
    \hline    
    Computation Model & DataFlow & DataFlow & DataFlow \\
    \hline
    Optimized for & Dense Matrix & SpMM & SpMV \\
    \hline   
    \end{tabular}
\end{table*}

In the preceding sections, we explored the TPU and its variants, the STPU and the FlexTPU. A comparison of these three architectures is provided in Table~\ref{tab:comparison}. From this comparison, it becomes clear that the choice of architecture depends on the nature of the workload. For dense matrix computations, the TPU is the preferred choice because of its optimization for such tasks. For sparse computations, the choice between STPU and FlexTPU depends on the specific operation: for SpMM, STPU is more suitable, while for SpMV, FlexTPU is the optimal choice. This understanding allows us to select the most efficient architecture for a given computational task.\\

The architectures we have discussed so far, including TPU, STPU, and FlexTPU, have primarily focused on optimizing the latency of matrix computations. However, because of the efficient utilization of their PEs, these architectures also offer energy savings.

There are certain situations where energy savings become the primary performance metric in our design. This is particularly true for technologies that rely on batteries or are used in remote locations, such as in space exploration or at the bottom of the ocean. In these scenarios, design choices must prioritize energy efficiency while still delivering good performance metrics.

In the following section, we will discuss an accelerator that has been specifically designed to optimize energy consumption, demonstrating how performance and efficiency can be balanced in hardware design.

\section{RipTide}

This section discusses RipTide, as presented in the paper "RipTide: A programmable, energy-minimal dataflow compiler and architecture” by Graham Gobieski et al. \cite{riptide}.

RipTide addresses the need for a system that is both highly programmable and extremely energy efficient.

Traditional computing architectures often struggle to balance these two aspects. On one hand, general-purpose processors, such as CPUs, offer high programmability, meaning they can handle a wide variety of tasks. However, they are not very energy efficient, especially for specific tasks like deep learning or signal processing.

On the other hand, specialized hardware accelerators, like GPUs or TPUs, can be very energy efficient for specific tasks but lack the broad programmability of CPUs. They are designed for specific types of computation and may not perform well outside of those tasks.

RipTide aims to bridge this gap by offering a solution that is both highly programmable, capable of handling a wide range of computations, and extremely energy efficient. This is particularly important for applications that require ultralow-power processing, such as emerging sensing applications.

RipTide achieves this through a unique combination of a co-designed compiler and a CGRA architecture.

\subsection{RipTide’s Architecture}
RipTide’s architecture is designed to achieve both high programmability and extreme energy efficiency by co-designing the compiler and CGRA, introducing control flow primitives, enforcing memory ordering and implementing control flow in NoC. The components of the architecture are described as follows.

\begin{enumerate}
  \item CGRA: The CGRA is an array of PEs connected by an on-chip network (NoC). The CGRA is programmed by mapping the dataflow of a computation to the array, i.e. assigning operations to PEs and configuring the NoC to route values between dependent operations.
  \item Control-Flow Operators: RipTide introduces control-flow primitives that support common programming idioms, deeply nested loops, and irregular memory accesses while minimizing energy overhead. These operators allow RipTide to support arbitrary control flow and memory access on the CGRA fabric.
  \item Memory Ordering: RipTide enforces memory ordering through careful analysis at compile time, eliminating the need for expensive tag-token matching on fabric. This means that it can handle complex memory access patterns without the need for costly hardware mechanisms.
  \item Network-on-Chip (NoC): RipTide implements control flow in the NoC to increase utilization and facilitate compilation. The NoC is responsible for routing values between dependent operations in the CGRA.
  \item Compiler: The RipTide compiler is implemented in LLVM and is designed to compile applications written in a high-level language with minimal energy and high performance. 
\end{enumerate}

A key feature of RipTide's architecture is its use of a hybrid computing model, which strikes a balance between energy efficiency and flexibility. The architecture is primarily designed to execute most operations on the CGRA, which is known for its high energy efficiency and reasonable flexibility.

However, there are certain operations, especially those that involve complex control flow and irregular memory access patterns, that are not ideally suited for the CGRA. In these situations, RipTide utilizes a more traditional von Neumann core. This core excels at managing these complex operations, offering a higher degree of flexibility.

It is important to note that, while the von Neumann core offers this increased flexibility, it does not achieve the energy efficiency of the CGRA. Therefore, RipTide’s architecture represents a strategic balance between flexibility and energy efficiency.\\

Our discussion thus far has centered on the design of accelerators using ASICs and reconfigurable architectures. However, it is worth noting that reconfigurable architectures can also be used in other applications, such as enhancing the performance of a datacenter while providing flexibility to adapt to evolving workloads. Microsoft currently has an active project in this area titled “Catapult” \cite{Microsoft}. We will discuss this in the next section.

\section{Catapult}

The paper titled “A Reconfigurable Fabric for Accelerating Large-Scale Datacenter Services” by Microsoft explores the challenges of enhancing computational capability, flexibility, power efficiency, and cost in datacenter workloads \cite{catapult}\cite{Microsoft}. As datacenter workloads evolve rapidly, continuous improvement in performance is required. As discussed previously, general-purpose architectures have limited efficiency in handling computational demands, while homogeneous accelerators lack the flexibility needed for rapidly evolving services in the datacenter.

Microsoft’s solution to this challenge is to use an FPGA to strike a balance between efficiency and flexibility. This implementation, known as the Catapult fabric, is a reconfigurable computing infrastructure designed for large-scale services.

The Catapult architecture integrates FPGAs directly into the datacenter server infrastructure. Each server hosts a Stratix V FPGA on a small daughtercard with a mezzanine connector for connectivity. The FPGA communicates with the FPGAs of other servers over a low-latency, high-bandwidth torus network. This architecture ensures that services requiring more than one FPGA can be mapped across FPGAs on multiple servers, thus improving elasticity and making efficient use of the reconfigurable fabric.

The architecture is divided into two partitions: a shell and a role. The shell provides reusable infrastructure such as PCIe, DRAM controllers, and interconnects, while the role is the application logic that can be reconfigured as needed. This design allows developers to focus on application logic without worrying about the low-level details of the infrastructure.

In addition, the Catapult Fabric has been deployed on a medium scale across 1,632 servers, demonstrating its ability to accelerate large-scale software services. It has shown significant improvements in performance and efficiency, achieving a 95\% improvement in throughput at each ranking server with an equivalent latency distribution. Alternatively, at the same throughput, it reduces tail latency by 29\%\cite{Microsoft}.

This work has pioneered the use of FPGAs in cloud computing, proving that FPGAs can deliver efficiency and performance without the cost, complexity, and risk of developing custom ASICs.

\section{Conclusion}

This paper emphasizes significant advances and challenges in computer architecture, focusing on the transition beyond multi-core processors toward domain-specific accelerators. 

As traditional scaling strategies face physical and power limitations, innovative solutions such as TPUs and their variants, including SparseTPU and FlexTPU, showcase how specialized architectures can efficiently handle complex computational tasks. These architectures offer customized performance improvements by optimizing for different types of parallelism and computational models. 

Meanwhile, innovations like RipTide demonstrate the potential of energy-efficient, flexible designs that bridge the gap between programmability and specialization. The Catapult fabric exemplifies the future of reconfigurable hardware, providing scalable solutions adaptable to evolving workloads in datacenters. 

Future research will continue to push the boundaries of architectural design, seeking an optimal balance between flexibility, performance, and energy efficiency to meet the growing demands of modern computing applications.

\end{document}